\documentstyle[psfig]{l-aa}

\def\kms{\,km\,s$^{-1}$}

\def\micron{\,$\mu$m}
\def\magarc{ mag~arcsec$^{-2}\,$}

\begin{document}

\thesaurus{20(11.19.2;11.19.6;13.09.1)}

\title{Near-infrared surface photometry of early-type spiral galaxies:
I. Bulge and disk decomposition\thanks{Based on
observations taken at TIRGO (Gornergrat, Switzerland). TIRGO
is operated by CAISMI--CNR, Arcetri, Firenze, Italy.}}

\author{G. Moriondo \inst{1} \and C. Giovanardi \inst{2}
\and L. K. Hunt \inst{3}}
\institute{ 
  Dipartimento di Astronomia e Scienza dello spazio, L. E. Fermi 5,
  I-50125 Firenze, Italy
\and
  Osservatorio Astrofisico di Arcetri, L. E. Fermi 5, I-50125 Firenze, Italy
\and
 C.A.I.S.M.I., L. E. Fermi 5, I-50125 Firenze, Italy}

\offprints{G.~Moriondo}
\date{Received 22 May 1997; accepted }

\maketitle
\markboth{Moriondo et al.: NIR photometry of early-type spirals. I}{Moriondo et
 al.: NIR photometry of early-type spirals. I}

\begin{abstract}
We present near-infrared (NIR) surface photometry of a sample of
14 early-type spirals with observed rotation curves.
In this first paper, we report the results of
two-dimensional parametric and non-parametric decompositions to separate
the bulge and disk components;
the parametric bulge is modeled with a generalized exponential law of integer index $n$,
and the disk with a simple exponential. 
We find that the derived bulge parameters, for a given galaxy,
vary systematically with the bulge shape index $n$. 
The mean early-type bulge has a best-fit $n$~=~2.6, and
80\% of the sample has best $n$ of 2 or 3.
Bulges are rarely spherical; the median bulge intrinsic ellipticity is 0.33.
The median early-type disk has $(J-K)_d$ more than 0.1~mag bluer 
than the bulge, and a NIR disk surface brightness 
more than 1 mag~arcsec$^{-2}$ brighter than later-type disks.
Our data are consistent with the well-established correlation of 
both bulge and disk surface brightness with physical
scale length, and we note that the location of bulges within this 
projection of the fundamental plane depends on their shape 
index $n$.
In agreement with previous work,
the ratios of bulge and disk scale lengths are consistent with
a constant value $r_e/r_d$~=~0.3; 
however, such value again depends on the bulge index $n$,
implying that claims for a scale-free Hubble sequence may be premature.
\end{abstract}


\section{Introduction}

Surface brightness distributions of external galaxies have been studied 
for many years but 
a reliable decomposition into structural components is
often difficult to obtain.
The reliability of models and techniques 
has been questioned by several authors 
(Kent \cite{kent:1986}; Schombert \& Bothun \cite{schombert};
Byun \& Freeman \cite{byun:freeman}), 
and there are several points that need further study and refinement.
First, brightness distributions have often been studied as
one-dimensional (1-D) radial profiles extracted by averaging along 
elliptical annuli; such ellipses
can deviate considerably from the actual isophotes, especially
in highly inclined systems with a luminous bulge.
The result provides therefore a distorted profile of the
surface brightness along the major axis. 
Second, the seeing must be properly taken into account
(c.f., Schombert \& Bothun \cite{schombert}),
especially when studying the central regions where brightness gradients are 
highest. 
Third, the choice among the parametric forms of 
brightness distribution (exponentials, Hubble and 
de Vaucouleurs laws, and so on) does not depend on any physical argument, but
only on their ability to fit the data. 
Finally, the effects of internal extinction should be taken into account.

The new near-infrared (NIR) panoramic
detectors have made imaging 
in the 1 to 2.5\micron\ regime rather straightforward with
sensitivity and accuracy comparable to those attainable at optical
wavelengths. 
The NIR bandpasses have been
advocated to be the ideal ones to study the characteristics
of the galactic backbones, that is the stellar populations which 
make up the mass distribution in a galaxy (e.g., Rix \& Rieke \cite{rix:rieke}).
The reason for this is twofold: first, the extinction is lower, by a factor
of ten between the $B$ and $K$ bandpasses; and second, the 
emission of old stellar populations peaks in the NIR.

In this context,
we have undertaken a program of NIR imaging 
of bright spiral galaxies with measured rotation curves.
These images are used to decompose the luminosity
distribution into bulge and disk components and then to analyze their
contribution to the observed rotation curves.
To overcome the methodological drawbacks mentioned above, 
we have developed a parametric technique to fit a two-dimensional (2-D)
bulge $+$ disk distribution to the entire image, which 
takes into account the effect of seeing. 
In addition, generalizing Kent's non-parametric approach, we have developed 
an iterative algorithm to deduce from the 2-D brightness distribution
the contributions of bulge and disk, again taking into account the seeing.
In this first paper, we report the results of the
structural decomposition for a restricted sample of early-type spirals;
a subsequent paper will describe the inferred mass distributions. 
Such systems appear well suited for studies of this kind,
mainly because of the relative smoothness of their disks and their (reputedly)
lower internal extinction. 
On the other hand, due to the low gas content,
rotation in Sa's is not measured as far out as for
later types, and hence they are not quite as effective 
for exploring dark matter properties. 
In several cases it has been
noted that the presence of a dominant bulge might severely
complicate the picture and the analysis (Kent \cite{kent:1988}); 
these are galaxies with luminous
bulges and slowly rising rotation curves, 
and some of them are also included in our sample.

\section{Observations and data reduction}

We selected 14 early-type spirals with observed rotation curves 
(Rubin et al. \cite{rubin}; Fillmore et al. \cite{fillmore}), 
and with diameter $\leq$ 8~arcmin. 
so that they could be imaged within our field of view.
The sample is listed in Table \ref{table:sample}\protect, 
objects with anomalously slowly rising rotation curves are marked with an 
asterisk.
Morphological type and coordinates are taken from NED\footnote{
The NASA/IPAC Extragalactic Database (NED) is operated by the Jet Propulsion
Laboratory, California Institute of Technology, under contract with the U.S.
National Aeronautics and Space Administration.}.
Three galaxies reside in the Virgo Cluster: NGC~4419,
4450 and 4698 (Binggeli et al. \cite{binggeli}). 
The remaining 11 
are members of loose groups (Huchra \& Geller \cite{huchra:geller};
Geller \& Huchra \cite{geller:huchra}; Rubin et al. \cite{rubin}; 
Garcia \cite{garcia}). 
Distances were computed 
according to the infall model in the Third Reference Catalog of Bright
Galaxies (RC3, de Vaucouleurs et al. \cite{devauc:rc3}),
and assuming an $H_\circ$~=~50\kms \,Mpc$^{-1}$.
For each group or cluster its mean redshift was used;
the adopted distance scale implies a Virgo distance modulus of 31.9 mag.
The position angle (Column 6) is obtained at the outer isophotes, measured 
East from North; the optical radius and
the $B$-band magnitude (Columns 7, 8) are from the RC3; 
Column 9 reports the bands in which each galaxy has been observed.

\begin{table*}
\begin{flushleft}
\caption{The galaxy sample}
\label{table:sample}\protect
\begin{tabular}{llllllllrrrcl} 
\noalign{\smallskip}
\hline
\noalign{\smallskip}
\multicolumn{1}{c}{Name} & \multicolumn{1}{c}{Type} & 
\multicolumn{3}{c}{R.A. (1950)} & \multicolumn{3}{c}{Dec. (1950)} & 
\multicolumn{1}{c}{Distance} & \multicolumn{1}{c}{Pos. angle} & 
\multicolumn{1}{c}{$R_{25}$} & $m_B$ & \multicolumn{1}{c}{Band} \\
     &      & h & m & s       & $^o$ & ' & ''   & (Mpc) & (deg)  & (arcsec) &  (mag) &  \\
\multicolumn{1}{c}{(1)} & \multicolumn{1}{c}{(2)} & \multicolumn{3}{c}{(3)} & 
\multicolumn{3}{c}{(4)} & \multicolumn{1}{r}{(5)} & \multicolumn{1}{r}{(6)} & 
\multicolumn{1}{r}{(7)} & \multicolumn{1}{c}{(8)} & \multicolumn{1}{c}{(9)} \\
\noalign{\smallskip}
\hline
\noalign{\smallskip}
N1024 & (R')SA(r)ab     & 02 & 36 & 30.4 & 10 & 37 & 56 & 69.2  &  155 & 117 & 13.08 & $K$   \\
N2639 & (R)SA(r)a:? Sy1 & 08 & 40 & 03.0 & 50 & 23 & 11 & 70.4  &  140 & 55  & 12.56 & $J,K$ \\
N2775 & SA(r)ab         & 09 & 07 & 41.0 & 07 & 14 & 35 & 27.1  &  155 & 128 & 11.03 & $J,K$ \\
N2841$^*$ & SA(r)b:         & 09 & 18 & 35.8 & 51 & 12 & 31 & 19.2  &  147 & 244 & 10.09 & $J,K$ \\
N3593 & SA(s)0/a:       & 11 & 11 & 59.2 & 13 & 05 & 28 & 17.2  &   92 & 157 & 11.86 & $J,K$ \\
N3898$^*$ & SA(s)ab         & 11 & 46 & 36.1 & 56 & 21 & 42 & 34.8  &  107 & 131 & 11.60 & $J,K$ \\
N4378 & (R)SA(s)a       & 12 & 22 & 44.8 & 05 & 12 & 06 & 51.4  &  167 & 87  & 12.63 & $J,K$ \\
N4419 & SB(s)a          & 12 & 24 & 24.6 & 15 & 19 & 24 & 23.9  &  133 & 99  & 12.08 & $J,K$ \\
N4450 & SA(s)ab         & 12 & 25 & 58.0 & 17 & 21 & 42 & 23.9  &   11 & 157 & 10.90 & $J,K$ \\
N4698 & SA(s)ab         & 12 & 45 & 51.3 & 08 & 45 & 35 & 23.9  &  168 & 119 & 11.46 & $J,K$ \\
N4845 & SA(s)ab         & 12 & 55 & 28.1 & 01 & 50 & 42 & 27.1  &   89 & 150 & 12.10 & $J,K$ \\
N5879$^*$ & SA(rs)bc:?      & 15 & 08 & 27.6 & 57 & 11 & 25 & 26.2  &    2 & 125 & 12.22 & $J,K$ \\
N6314 & SA(s)a: sp      & 17 & 10 & 33.1 & 23 & 19 & 43 & 139.2 &  177 & 43  & 13.80 & $J,K$ \\
IC724 & Sa              & 11 & 41 & 00.1 & 09 & 13 & 10 & 125.6 &   60 & 70  & 13.40 & $J,K$ \\ 
\noalign{\smallskip}
\hline
      &         &   &   &            &          &       &      &    &   & \\
\end{tabular} 
\end{flushleft}
\end{table*}

Images were acquired at the Gornergrat Infrared Telescope 
(TIRGO, 1.5 m, f/20) 
in the $J$ (1.2\micron ) and $K$ (2.2\micron ) bands, 
using the Arcetri NIR camera ARNICA (Lisi et al. \cite{lisi}), 
equipped with a NICMOS~3 detector. 
The field of view of the camera is 4$\times$4~arcmin with a plate
scale of 0.97 arcsec\, pixel$^{-1}$.
Galaxies were observed by spending half of the total integration time
($\sim 15$ min) on source, and half on the sky,
alternating sky and source frames. 
The seeing FWHM was typically $2''$,
and sky magnitudes were roughly 15.5 and 12.5\magarc in $J$ and $K$,
respectively.
Flat fields were obtained from sky frames;
the final images are typically flat to 0.3--0.5\% in the $J$ band,
and to 0.1\% or better in the $K$ band, giving  
(1\,$\sigma$) limiting magnitudes of 
21.5 and 20.5\magarc in $J$ and $K$ respectively. 
All image reduction was performed with IRAF and the STSDAS
packages\footnote{IRAF is the Image Analysis and Reduction Facility
made available to the astronomical community by the National Optical
Astronomy Observatories, which are operated by AURA, Inc., under
contract with the U.S. National Science Foundation.
STSDAS is distributed by the Space
Telescope Science Institute, which is operated by the Association of
Universities for Research in Astronomy (AURA), Inc., under NASA contract
NAS 5--26555.}.
Standard stars were selected from the
UKIRT Faint Standard List (Casali \& Hawarden \cite{casali}). 
Zero points were derived for each night of observations after
correcting for atmospheric extinction using mean TIRGO 
values (0.11 and 0.07 mag airmass$^{-1}$
in $J$ and $K$ respectively; Hunt et al. \cite{hunt:calamai}). 
Images were then corrected for extinction within our Galaxy and
for redshift.
Galactic extinction was determined according to the
Burstein--Heiles values given in RC3, using
the extinction curve given in Cardelli et al. (\cite{cardelli}); 
K-corrections for redshift
were calculated from the coefficients given by Frogel et al. (\cite{frogel}).

Color images were made by combining the $J$- and $K$-band magnitude images
after first registering them to a fraction of a pixel. 
When the seeing differed significantly between the two bands,
the higher-resolution image was degraded to the level of the lower
resolution one.
The $K$-band images and $J-K$ color images are shown in the top panels
of Fig. \ref{figure:all}.

For five of the sample galaxies
we can compare the $K$-band major-axis profiles with low-resolution 
(28'' aperture)
profiles obtained with a single-element photometer
(Giovanardi \& Hunt \cite{giova:1996});
the data are shown as filled circles in Fig. \ref{figure:all}.
As can be seen, the agreement is quite good
in spite of the pointing uncertainties in the photometer profiles.

\section{Quantitative morphology}

\subsection{\protect\label{pd} The parametric decomposition}

For the 2-D parametric fits, 
we have modeled the surface brightness distribution
with a generalized exponential bulge 
(S\`ersic~\cite{sersic}, Sparks \cite{sparks}): 
\begin{eqnarray}
\label{bulge_b}\protect
\lefteqn{ I_b\,(x,y)  \, = } \nonumber \\
 & & I_e \, \exp
\left\{-\alpha_n \left[\left(\frac{1}{r_e}\sqrt{x^2+\frac{y^2}
{(1-\epsilon_b)^2}}\, \right)^{1/n}-1\right]\right\},
\end{eqnarray}
plus an exponential thin disk: 
\begin{equation}
\label{disk_b}\protect
I_d\,(x,y)\ =\ I_d(0)\exp\left[-\frac{1}{r_d}\sqrt{x^2+
\frac{y^2}{\cos^2 \,i}}\ \right] \; .
\end{equation}
$I_e$ and $r_e$ are effective (half-light) values, 
$\epsilon_b$ is the apparent bulge ellipticity, 
$\alpha_n$ is a constant relating 
the effective brightness and radius to the exponential values (see Appendix).
$x$ and $y$ are in arbitrary units, with origin
at the galaxy's center, and 
$x$ along the major axis. 
The bulge is assumed to be an oblate rotational ellipsoid, 
coaxial with the disk, and its 
apparent eccentricity $e_b$ is related to
the intrinsic eccentricity $e_b'$ by:
\begin{equation}
\label{ecce}\protect
e_b = e_b'\sin i \;\; .
\end{equation}

The structural parameters have been determined 
by fitting the photometric data to the model, convolved with a circular 
Gaussian seeing disk of appropriate FWHM, using a $\chi^{2}$ minimization. 
The fitted parameters are:
the two surface brightnesses $I_e$ and $I_d(0)$ (or 
$\mu_e$ and $\mu_d$ when given in magnitudes), 
the two scale lengths $r_e$ and $r_d$, the bulge ellipticity
$\epsilon_b$, and the system inclination $i$. 
Because of the difficulty inherent in estimating a luminosity distribution
of unknown form (bulge) which is partially embedded in another (disk), 
we did not attempt to fit the bulge index $n$ explicitly (c.f. 
Andredakis et al. \cite{andredakis:peletier}).
Instead, for each galaxy the bulge was modeled
with four different values of the exponent $n$:
1, 2, 3, and 4 ($n\,=\,4$ for a de Vaucouleurs bulge). 
We then assessed the ``quality'' of the results for each set of 2-D fits,
and determined the value of $n$ that gave the best all-round fit, thus
assigning, in effect, a shape index to the bulge.
Three quality indicators were considered in the assessment: 
the successful convergence of the process,
the value of $\chi^2$, and the mean absolute residual.

When both $J$ and $K$ data were available, 
tests showed that bulge and disk decompositions and
colors of the components were more stable 
when the two bands were fitted simultaneously.
We also found no evidence for systematic scale length 
changes with wavelength of either component (c.f., Evans \cite{evans}).
Hence, when two bands were available, we performed
a single fit keeping scale lengths, inclination, and bulge ellipticity
the same for both bands, and letting the surface brightnesses of both
components vary independently, but with constant color indexes 
$(J-K)_b$ and $(J-K)_d$~.

The decomposition technique was extensively tested on
a set of synthetic maps of axisymmetric bulge$+$disk distributions, 
covering a range of values for $i$, $\epsilon_b$, and B/D ratio similar to ours.
We adopted an exponential thin disk, and generalized exponential laws for
the bulge. 
An appropriate amount of noise was added to
simulate the typical signal-to-noise levels we achieve, and finally
the models were convolved with a range of PSF's to
simulate the effects of seeing.
The 2-D decomposition method was always able to recover the true parameter set,
within the estimated errors, independently of the starting point in
parameter space.

We also performed a 1-D fit to the radial
profile, obtained by averaging along elliptical annuli;
center and position angle of the annuli were held fixed, while the 
ellipticity was allowed to vary with semi-major axis.
In the 1-D case, the convolution of the model with the PSF requires 
$i$ and $\epsilon_b$ to be assigned a priori:
as customary in 1-D techniques, $\epsilon_b$ was fixed at 0 and $i$
was determined from the outer $J$-band isophotes.
The best-fit bulge index $n$ was determined in the same way as for
the 2-D models.
1- and 2-D techniques are compared in $\S4.1$.

\begin{figure*}
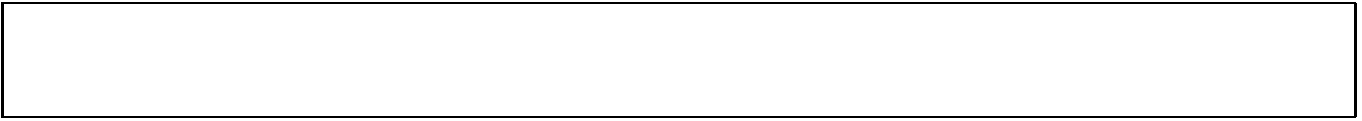

\picplace{1.5cm}
\caption[]{$K$-band images, $J-K$ color images, parametric, and 
non-parametric decompositions.
The upper left panel shows the $K$-band map;
contours range from 21 to 14\magarc in unit steps.
Images are oriented with North up and East left, and
offsets are given in arcsec.
The upper right panel shows the $J-K$ color map, with the color
table ranging from 0.5 (black) to 1.3 (white).
The lower left panels show the results of the parametric decomposition
with the $K$-band elliptically-averaged radial profiles shown as filled
triangles, and the major-axis cut shown as a solid line.
The parametric bulge and disk are shown as dot-dashed and dashed
lines, respectively, and their sum as a dotted line.
28$''$ aperture profiles (Giovanardi \& Hunt \cite{giova:1996})
are shown as large filled circles, if available.
The remaining lower left panels show the elliptically-averaged profiles
in $J-K$, and $r-K$ with the $r$-band taken from Kent (\cite{kent:1988}),
if available.
Also shown in the $J-K$ plots are
the fitted color profile (dotted line), and the fitted colors of
bulge (dot-dashed line) and disk (dashed line).
The lower right panel shows the results of the non-parametric
decompositions, with the same coding as for the parametric decompositions.
}
\label{figure:all}
\end{figure*}

The lower left panels in Fig. \ref{figure:all} show the results of the 
2-D decompositions in terms of cuts along the semi-major axis ($K$ band). 
Table \ref{table:parameters} gives the best-fit parameters, including
the shape index $n$:
the first line reports the values from the 2-D fit,
the second reports the 1-D ones, and
the third gives the values from the non-parametric decompositions
($\S$~\ref{npd}). 
Bulge and disk
surface brightnesses are corrected to face-on values assuming optical 
transparency with  C\,=\,1: 
\begin{eqnarray}
\label{eqn:transparency}\protect
\mu_d^c & = & \mu_d -\ 2.5\ C\ \log\, (\cos\, i) \quad, \\
\mu_e^c & = & \mu_e -\ 2.5\ C\ \log\, (1-\epsilon_b)\quad\nonumber , 
\end{eqnarray}
and are denoted by $\mu_e^c(K)$ and $\mu_d^c(K)$ respectively.
Table~\ref{table:luminosities} lists the derived luminosities
and bulge-to-disk ratios (B/D);
the first line reports the 2-D parametric fit values, and the second
gives the non-parametric values (see Sect.~\ref{npd}).
The 1-D fixed-position-angle profiles together with Tables 
\ref{table:parameters} and \ref{table:luminosities} are available in
electronic form on the ftp node: 
{\tt sisifo.arcetri.astro.it} in directory {\tt ~ftp/pub/nir}.

\begin{table*}
\caption[]{Decomposition results: structural parameters}
\label{table:parameters}\protect
\vspace {1.5cm}
\end{table*}

\subsection{\protect\label{npd} The non-parametric decomposition}

Kent (\cite{kent:1986}) was the first to introduce a non-parametric (np) method
to separate bulge and disk; 
recently this has been extended to the profile analysis of 2-D
images by Andredakis et al. (\cite{andredakis:peletier}).
We propose here a new, completely 2-D technique.
Assuming the apparent ellipticities of the two components, $\epsilon_b$ 
and $\epsilon_d=1-\cos\, i$, are known,
let us consider the average surface brightness
of an annulus of ellipticity, say, $\epsilon_d$.
For a sufficiently narrow annulus, $A$, the disk contribution 
will be $I_d(a)$, where 
$a$ is the semi-major axis.
The total average brightness will be: 
\begin{equation}
\label{int_an}\protect
<I(a)>_d=I_d(a)+\frac{1}{S_A}\int_{A}I_b \,dS
\end{equation}
where $S_A$ is the annular area.
An analogous equation can be written for an annulus of 
ellipticity $\epsilon_b$. 
The integral in 
Eq. (\ref{int_an}\protect) can be written as: 
\begin{equation}
\label{int_form}\protect
\frac{2S_A}{\pi}\int_0^{\pi/2} I_b \left( a\sqrt{\cos^2\theta+
\frac{\rho^2_d}{\rho^2_b}\sin^2\theta}\ \right) \,d\theta 
\end{equation}
where $\rho_b$ 
and $\rho_d$ are the axial ratios of the components.
We then obtain for any $a$,
a system of integral equations:
\begin{equation}
\label{boh}\protect
\left\{ \begin{array}{lll}
       {\displaystyle I_b(a)} & {\displaystyle= \;\; <I(a)>_b }& \\
	\\
       & {\displaystyle -\frac{2}{\pi}\int_0^{\pi/2} I_d\left(a
       \sqrt{\cos^2\theta+\frac{\rho^2_b}{\rho^2_d}\sin^2\theta}
       \ \right)\, d\theta} & \\
       \\
       \\
       {\displaystyle I_d(a)} & {\displaystyle = \; <I(a)>_d } & \\
       & {\displaystyle -\frac{2}{\pi}\int_0^{\pi/2} I_b\left(a
       \sqrt{\cos^2\theta+\frac{\rho^2_d}{\rho^2_b}\sin^2\theta}
       \ \right)\, d\theta} & 
       \end{array}
\right.
\end{equation} 
To solve the system we first compute $<I(a)>_b$ and $<I(a)>_d$ for
two sets of annuli of ellipticities $\epsilon_b$ and $\epsilon_d$, respectively,
each set sampling the entire light distribution. 
The two equations can then be solved iteratively for $I_b$ and $I_d$
at each $a$, starting with 
$<I>_b$ and $<I>_d$ as initial guesses 
to evaluate the integrals. 

When defining the two sets of annuli, 
a proper sampling of the semi-major axis is of crucial importance. 
A reliable estimate of the integrals in
Eq.~\ref{boh}\protect ~requires that each annulus of either
set intersects with a sufficient number of annuli of the other set.
If we constrain this number to be the same for every ellipse, the 
sampling 
is linear in $\log a$, i.e. the annuli become wider with increasing radius.

Results of the np decomposition technique are given in the
third line of Table \ref{table:parameters},
and in the second line of Table \ref{table:luminosities}.
The non-parametric parameters are defined as follows. The bulge 
$r_e$ is the semimajor axis of the ellipse enclosing half light; 
$\mu_e$ is the bulge surface brightness at $r_e$; $r_d$ is the
semimajor axis of the ellipse enclosing 0.264 of the disk light,
as in the exponential case. Since the central disk is not
tightly constrained by the np fits, $\mu_d$ is the central brightness
of an exponential disk having scale length $r_d$ and the same
total luminosity as the np disk. The np colors of the components are
derived by the total component luminosities in the two bands.

\subsubsection{Testing of non-parametric decomposition}

The np decomposition algorithm was tested 
on the same set of synthetic maps 
used for the parametric tests.
Since the method applies only to components with different apparent ellipticity,
we only tested models with $\epsilon_b$ 
at least 10\% smaller than $\epsilon_d$. 
Under these conditions, to within the noise level, 
the method always proved to be able to recover the true
bulge and disk distributions given the correct values 
of the ellipticities. 

When errors on the ellipticities are introduced,
the shape of the resulting components is affected in a systematic way:
the value of $\epsilon_b$ affects mainly the inner part of the disk,
while $\epsilon_d$ determines the shape of the outer bulge. 
In particular, if $\epsilon_b$ is underestimated the inner disk 
is too steep, whereas if the bulge is too
elliptical the inner disk develops a hole which gets deeper as $\epsilon_b$
increases. 
If $\epsilon_d$ is too small the outer bulge is too steep and viceversa.
Although qualitatively these are expected trends, they
are difficult to quantify in practice, 
since they depend on the 
shapes of the distributions, on the ellipticities,
and on the seeing. 
Especially  when the two components have similar 
ellipticities, even small errors on $\epsilon_b$ and $\epsilon_d$ ($<0.1$) 
perceptibly affect parameters such as the 
luminosity of the components or the B/D ratio, as well as the 
shape of the profiles.

\subsubsection{Seeing corrections}

Simulations show that, if no correction is applied, 
seeing significantly affects np decompositions. 
The effects are stronger in the inner parts, 
the main feature being a central 
hole in the disk, whose extension is proportional to the seeing 
width. 
In fact, the seeing makes 
inner isophotes rounder, causing the component of higher ellipticity 
to be depressed. The disk profile, however, can be 
corrected by extrapolating the outer points to the center.

To recover the intrinsic
bulge profile, we introduced a
second iterative algorithm which proved to be quite effective on
synthetic maps. 
The effect of seeing is accounted for by
defining a correction coefficient $k_b(a)$, for each
point of the bulge profile, as the ratio of the convolved 
bulge distribution and an estimate of the true one. 
Each decomposition is performed using $k_b(a)$
to degrade the assumed true distribution.
After decomposition, the coefficients
are recomputed by convolving the extracted bulge profile with
the PSF, and a new decomposition is performed until 
both the coefficients and the profiles converge.
The initial estimate of $k_b(a)$ is provided by the
convolution of the bulge profile extracted without any correction.

\subsubsection{Estimate of the ellipticities}

To determine the apparent ellipticity of the components, 
we extracted from each image a radial 
profile of the ellipticity by fitting 
isophotes of variable ellipticity and fixed position angle.
We then selected  
the lowest value beyond one seeing
disk from the center as $\epsilon_b$, 
and the average value in the outer parts as $\epsilon_d$
(see also Andredakis et al. \cite{andredakis:peletier}). 
The values for the $J$ and $K$ bands were then averaged
to obtain a single value of $\epsilon_b$ and $\epsilon_d$ for each 
galaxy.
The accuracy of these estimates is difficult to assess: 
bright bulges can affect the 
isophotes even at large radii, leading to 
an underestimate of $\epsilon_d$, while bars 
and spiral arms may often lead to overestimates. 
In turn, $\epsilon_b$ can be easily underestimated because of the
seeing.
From simulations, it seems
that a reasonable estimate for the uncertainty in $\epsilon_d$ is 
about 0.1; it is certainly worse in $\epsilon_b$, 
0.2 or more in the worst cases.

We note that comparing the observed image to the extracted
bulge$+$disk distribution, for example in terms of residuals, 
is of little use to test the ellipticities, 
since it is often possible to find excellent agreement 
for a wide range of $\epsilon_b$ and $\epsilon_d$.
A more stringent constraint is provided by the {\it color} profiles of bulge
and disk, which turn out to be quite sensitive to the values of $\epsilon_b$ and
$\epsilon_d$. This has ultimately allowed us to verify the ellipticities 
a posteriori (in particular $\epsilon_b$) by
letting them vary within the estimated errors
and checking the plausibility of the resulting color profiles. 
The final accuracy is estimated to be better than 0.05.

\begin{table}
\caption{Decomposition results: $K$-band integrated luminosities and bulge-to-disk ratios}
\label{table:luminosities}
\vspace {1.5cm}
\end{table}

\begin{figure*}
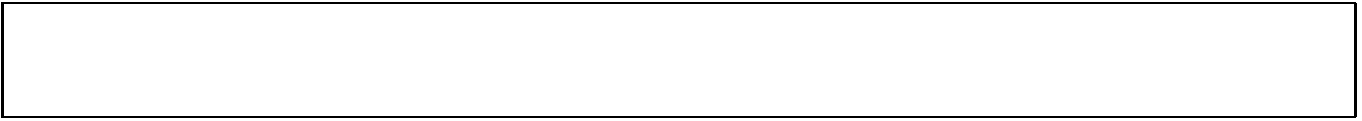

\picplace{1.5cm}
\caption{Bulge and disk fitted parameter variation with $n$.
Bulge parameters are shown in the left panels: from top to bottom, 
absolute $K$-magnitude $M_b(K)$, effective surface brightness $\mu_e^c(K)$,
and effective scale length $r_e$.
Analogous disk parameters are shown in the right panels.
The curves show, for a given galaxy, fit parameters
obtained with $n$ going from 1 to 4.
$\mu_e^c(K)$ and $\mu_d^c(K)$ are corrected to face-on
and in units of \magarc . Scale lengths are in arcsec.}
\label{figure:n}
\end{figure*}

\section{Results}

\subsection{Accuracy and comparison of the different decomposition methods
\protect\label{results:comparison}}

As Fig.~\ref{figure:all} clearly shows, the elliptically-averaged profiles 
reproduce quite well the major-axis cut in almost all cases. 
In fact,
our sample contains only moderately inclined systems ($i < 75^{\circ}$), 
whose isophotes are reasonably elliptical.
However, a comparison between the 1-D and 2-D results from 
Table~\ref{table:parameters} 
shows that the 1-D decomposition is always less
accurate, and sometimes inconsistent with the 2-D one. 
The inferior quality of the 1-D fits is 
attested to both by the larger errors on the parameters and 
by the larger values of $\chi^2$. 
A comparison between 1-D and 2-D results for the galaxies with significant
non-axisymmetric structure suggests that 
2-D fits manage to reproduce the whole distribution,
whereas 1-D decompositions tend to match closely only the 
inner part of the profile, i.e. the less noisy one 
(see also Byun \& Freeman \cite{byun:freeman}).
Because of their superiority, when discussing parametric models,
we will consider only the results of 2-D decompositions.

No systematic differences are found between the various fitted parameters:
the 1-D bulge and disk scale lengths are, to within the scatter, 
the same as those of the 2-D fits, and the fitted inclination of the 2-D model
corresponds, on average, to the fixed value determined from the outer
isophotes (and used in the 1-D fits).
However, disks are definitely better constrained than bulges:
the rms difference of the 1- and 2-D 
$\mu_d$'s (0.6~$K$-mag)
is half that for $\mu_e$ (1.1~$K$-mag), and
the dispersion in the ratio of 1- and 2-D $r_e$'s is 80\%,
while that for the $r_d$'s is four times lower (21\%),
comparable to the discrepancies among disk scale lengths from 
different authors found by Knapen \& van der Kruit (\cite{knapen:vanderkruit}).

We have compared the results of our non-parametric (np) decompositions
with the 2-D parametric ones, and with those of Kent (\cite{kent:1988}).
There is no systematic difference between our 2-D and np bulge and disk
luminosities, as the mean difference 
is $-0.4\,\pm\,0.6$~mag for $M_b$, and $-0.1\,\pm\,0.4$ for the disk. 
Again, disk parameters are more consistent than those of the bulge.
Parametric ellipticities are not significantly different from np ones:
the mean ratio of the np $\epsilon_b$'s and those
fitted by the 2-D parametric method is 1.1~$\pm$~0.5.
System inclinations agree very well, with a mean ratio of $0.97\,\pm\,0.09$. 
Also, the bulge ellipticities determined by Kent (\cite{kent:1988}) are about
the same as those used in our np decomposition, but with large scatter;
the inclinations are in good agreement, with a mean ratio
of $1.01\,\pm\,0.09$.

\subsection{Dependence on $n$ of 
parametric components \protect\label{results:n}}

In order to compare our results with those of other authors, 
we have investigated trends with $n$ of the bulge and disk fitted parameters.
Figure \ref{figure:n} shows the systematic variation in these, 
for a given galaxy, as a function of $n$.
It can be seen from the figure that the derived bulge parameters 
strongly depend on the form of the fitting function.
The same bulge, when fitted with small $n$, appears to be ``denser''
(that is to say brighter $\mu_e$), more compact (smaller $r_e$),
and less luminous than when fitted with large $n$.
Quantitatively, the changes are dramatic as $n$ goes from 1 to 4 with
a mean change in $\mu_e$ of 3~mag~arcsec$^{-2}$, and in 
scale length of roughly a factor of 3.
Moreover, the dispersion in the fitted values also increases with $n$;
the spread of $\log\,r_e$
at $n$~=~4 is 1.5 times larger than that at $n$~=~1, and
the spread of $\mu_e$ is more than twice as large. 

The derived disk parameters also change with the $n$ of the bulge.
$\mu_d$ tends to be fainter for bulge $n$ larger, and,
as for the bulge, the dispersion in $\mu_d$ increases with $n$.
The only parameter that is stable with $n$ is 
the disk scale length $r_d$, although its dispersion does increase slightly
with $n$.

We conclude that, at least statistically, bulge structural parameters
are strongly influenced by the form of the function used to derive them.
Not only are the parameters themselves altered by constraining the form of
the bulge, but also the dispersion in the parameters is changed.
Independently of the best-fit $n$, requiring a de Vaucouleurs law to fit
the bulge yields more tenuous, extended, and luminous spheroids, together
with wider distributions of the parameters (larger dispersion),
than does using a simple exponential.

\subsection{Bulges \protect\label{results:bulges}}

In terms of the ``quality indicators'' mentioned in Sect. \ref{pd},
$n$~=~3 gave superior results for the majority of galaxies, while
$n$~=~2 was the second-best choice.
In only two cases did $n$~=~4 give the highest quality fit, and in
one case (NGC~3593) $n$~=~1. 
These values are in general agreement with the trend noted by
Andredakis et al. (\cite{andredakis:peletier}) who fitted the $n$ of 
non-parametric bulge profiles and found that 
early-type spirals tend to have bulges with $n\ \sim \ 2-3$.
We also note that, in our experience, it is usually difficult 
to determine the best $n$ with a precision much better than 1.

\begin{figure*}
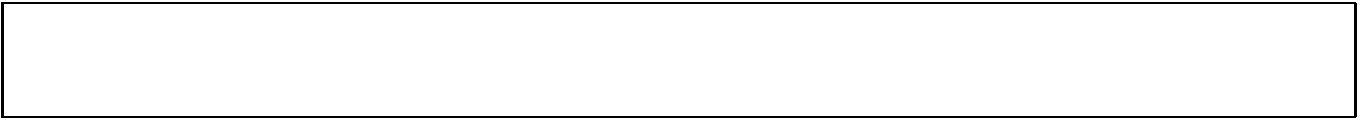

\picplace{1.5cm}
\caption[]{Distributions of best-$n$ bulge and disk parameters.
Surface brightnesses are corrected to face-on
and in units of \magarc . Scale lengths are in kpc.}
\label{figure:histogram}
\end{figure*}

The distribution of the best-$n$ bulge parameters 
is illustrated in the left panels of Fig.~\ref{figure:histogram}.
The median bulge parameters are $\mu_e^c(K)=16.8$\magarc , and
$r_e=1.6$ kpc, as reported in Table~\ref{table:averages}. 
Our bulges are more tenuous, larger, and more luminous than those 
of similar type in de Jong (\cite{dejong:3}), 
as expected given the fixed $n$~=~1 used by him.
The median {\it apparent} $\epsilon_b$ is 0.24 from parametric 
decompositions and 0.25 in np decompositions.
This translates into a median {\it intrinsic} ellipticity of 0.36 or 0.33 
from np values.
Notably, 0.33 is also the commonest intrinsic ellipticity found in elliptical
galaxies, if they are assumed to be rotational ellipsoids, either oblate
or prolate (Mihalas \& Binney \cite{mihalas}).
In any case, bulges are rarely spherical and the results of studies 
assuming so should be treated with caution.

\begin{table}
\begin{flushleft}
\caption[]{Statistics of parametric components }
\label{table:averages}\protect
\begin{tabular}{llllc} 
\noalign{\smallskip}
\hline
\noalign{\smallskip}
& & \multicolumn{1}{c}{Mean ($\sigma$)} & 
\multicolumn{1}{c}{Median} & Range \\ 
\noalign{\smallskip}
\hline
\noalign{\smallskip}
\noalign{\smallskip}
 & $n$  & 2.6 (0.8) & 3 & 1 -- 4 \\
 & $r_e$ & 1.9 (1.6) & 1.6 & 0.4 -- 6.4 \\ 
Bulge & $\mu_e^c(K)$ & 16.7 (1.1) & 16.8 & 15.1 -- 18.9 \\ 
 & $\epsilon_b$ & 0.24 (0.13) & 0.23 & 0.04 -- 0.50 \\ 
 & $(J-K)_b$ & 1.13 (0.27) & 1.06 & 0.72 -- 1.55 \\
 & $M_b(K$) & -23.9 (1.0) & -23.8 & -25.3 -- -22.2 \\
\noalign{\smallskip}
\hline
\noalign{\smallskip}
 & $r_d$ & 4.9 (2.8) & 4.6 & 1.40 -- 11.4 \\
Disk  & $\mu_d^c(K)$ & 17.3 (0.78) & 17.1 &16.1 -- 18.9 \\
 & $(J-K)_d$ & 0.96 (0.08) & 0.94 & 0.85 -- 1.11 \\
 & $M_d(K$) & -24.4 (0.8) & -24.3 & -25.4 -- -22.5 \\ 
\noalign{\smallskip}
\noalign{\smallskip}
\hline
\end{tabular}
\end{flushleft}
\end{table}

\begin{figure*}
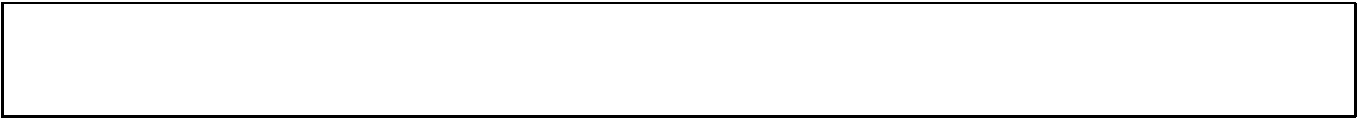

\picplace{1.5cm}
\caption[]{Scatter plots of bulge and disk $\mu^c(K)$ vs. $\log r$.
Scale lengths are in kpc.
The left panels show bulge parameters, and the right those of disks.
The upper panels show points for all values of $n$, with
$n$~=~1, 2, 3, 4 shown by $\times$'s, open circles, filled triangles,
and filled squares, respectively.
The lower panels show only the best-$n$ points, with $n$ denoted
as in the upper panel, together with the non-parametric results
shown as crosses.
In the lower panels, $<\mu^c(K)>_e$ is the
average surface brightness within the effective isophote; 
note that also for disks
the abscissa is $r_e$.
In the lower panels,
the dashed line shows the regression found by Andredakis et al.
(\cite{andredakis:peletier}) converted to our distance scale
and to the $K-$band according to their precepts. 
The inclined error bars represent the average shift
related to an
uncertainty of $\pm 1$ in the bulge index $n$. 
}
\label{figure:fp}
\end{figure*}

Correlations between (average) surface brightness and scale length have been
found for spiral bulges (Kent \cite{kent:1985}; Kodaira et al. 
\cite{kodaira:watanabe}) and
for ellipticals (Kormendy \cite{kormendy}; Hoessel \& Schneider \cite{hoessel};
Djorgovski \& Davis \cite{djor:davis}),
and these two observables
constitute an almost face-on view of the ``fundamental plane'' (FP)
(e.g., Kormendy \& Djorgovski \cite{kormendy:djor}, and references
therein).
The left panels of 
Fig. \ref{figure:fp} show scatter plots of bulge $\mu_e^c(K)$ 
and $<\mu^c(K)>_e$ vs. $r_e$,  $<\mu^c(K)>_e$ is the average surface brightness 
within the half-light isophote  commonly used in FP studies.
The upper panel shows results for all values of $n$, 
and the lower one only best-$n$ values.
Our results are consistent with the slope within the FP 
found for bulges by Andredakis et al. (\cite{andredakis:peletier}), 
shown as a dashed line in the figure. Although we have transformed 
their regression line to our distance scale and to the $K$-band 
according to their precepts, we note a slight offset: our best-$n$
bulges are generally dimmer ($\sim 0.5$ mag), for a given $r_e$, than theirs.

Figure \ref{figure:fp} gives further insights as to what happens when the 
form of the bulge is constrained to one value of $n$. 
As discussed in the previous section,
when for example $n$ is 4, the resulting bulge parameters
lie in the tenuous, extended portion of the parameter space;
when $n$ is 1, resulting bulges are smaller and denser
(see also Fig. \ref{figure:n}).
From Fig. \ref{figure:fp},
it appears that different $n$ values, that is to say different bulge shapes,
occupy different regions of the FP, such a behavior is evident in both
the upper- and lower-left graphs.
An analysis of the correlation coefficients 
shows that while the best-$n$ set of points is
significantly correlated, as is the global set of points for all values of $n$,
each individual set with fixed $n$ is not.
Moreover, the {\it slopes} of each fixed $n$ group increase with $n$,
although even the slope of $n$~=~4 is not as large as the global one.
The appearance of the top-left graph is determined mainly
by a "geometrical" effect, that is a constant luminosity relation
although, see Fig. \ref{figure:n}, higher $n$'s produce slightly
more luminous bulges. Independently of the details, it is clear that
the large scatter related to the uncertainty on the decomposition
has a high incidence on the position of a bulge within the FP, 
as shown by the error bar in the lower-left panel.

It has been suggested that
residuals relative to the FP are correlated with shape parameters
(Hjorth \& Madsen \cite{hjorth}; Prugniel \& Simien \cite{prugniel}).
Although this projection of the FP is not appropriate for such considerations,
the lower-left graph (bulge best $n$) in Fig. \ref{figure:fp}
suggests that even the distribution within the FP 
is at least partially generated by 
form variations. 
The distribution  of the bulge np parameters does not reveal any
dramatically different behavior; if anything, the distribution is 
tighter and situated in the low-$n$ region of the plot. 

\subsubsection{Bulge colors \protect\label{results:bulges.colors}}

The median $(J-K)_b$ of 1.06 (1.04 for the np decomposition)
agrees with the colors measured by Giovanardi \& Hunt (\cite{giova:1996}), and 
is redder by about 0.1--0.2~mag than those measured in later types
(Frogel \cite{frogel:1985}; Giovanardi \& Hunt \cite{giova:1988}).  
The scatter is large, 0.3~mag, with some bulges having
$J-K$ as high as 1.5 (NGC~4845).
We find that $(J-K)_b$ correlates 
(98\% significance) with $\mu_e^c(K)$, in the sense that redder colors are
associated with ``denser'' bulges.
In contrast,
$(J-K)_b$ is independent of $M_b(K)$, and of total galaxy luminosity.

The four objects (NGC 3593, 4419, 4845, and IC 724) with
$(J-K)_b\, >\,1.3$
also show red extended circumnuclear structure in the color images of 
Fig. \ref{figure:all},
inflections or bumps in their surface brightness profiles, 
and red gradients in the inner color profiles.
Such features have been observed in starburst galaxies 
(Hunt et al. \cite{hunt}),
and we would argue that the red $J-K$ bulge color is revealing
star formation in progress. 
The most clear-cut case is NGC~3593 which, besides
a high mid-infrared 12\micron\ surface brightness
(Soifer et al. \cite{soifer}) and 
high molecular gas content (Sage \cite{sage}),
hosts two counterrotating stellar disks
and a disk of ionized gas (Bertola et al. \cite{bertola}).
Moreover, $H\alpha$ images reveal an HII-region ring
(Pogge \& Eskridge \cite{pogge})
whose structure closely resembles that seen in our $J-K$ image.
NGC~4419, the only barred galaxy in our sample,
is a LINER (Huchra \& Burg \cite{huchra})
with mid-infrared properties (Soifer et al. \cite{soifer};
Devereux \cite{devereux}) and CO content (Young et al. \cite{young})
typical of starbursts.
NGC~4845 was defined as a starburst by David et al. (\cite{david})
on the basis of its FIR-to-blue luminosity ratio and X-ray excess.
IC~724, one of the most distant in our sample,
harbors more than 10$^9$\,M$_\odot$ of HI (Eder et al. \cite{eder}), but
we have found no evidence in the literature for star formation activity.
The $J-K$ image may be just revealing a normal bulge, partially obscured by a
dusty disk.

\subsection{Disks \protect\label{results:disks}}

\begin{figure*}
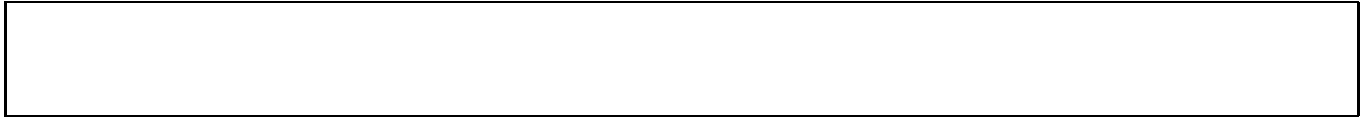

\picplace{1.5cm}
\caption[]{Disk scale lengths, ratio of bulge and disk scale lengths,
and B/D vs. optical isophotal radius $R_{25}$.
Best-$n$ values are shown in the left panels, and
values for all $n$ are shown on the right; symbols are as in 
Fig.~\ref{figure:fp}.
The upper panels show $\log r_d$ vs $\log R_{25}$;
the dotted line in the upper left panel illustrates a linear relationship
between $r_d$ and $R_{25}$ with a mean ratio $r_d/R_{25}$ of 0.24.
The middle panels show the ratio $\log r_e/r_d$ vs $\log R_{25}$;
the dotted line on the left gives the best-$n$ median
constant value $r_e/r_d$~=~0.3,
while the pair of lines on the right show the $n$~=~1 median $r_e/r_d$~=~0.2,
and the $n$~=~4 median $r_e/r_d$~=~0.7.
The lower panels show $\log(B/D)$ vs $\log R_{25}$.}
\label{figure:sl}
\end{figure*}

The distribution of the disk parameters is shown in 
the right panels of Fig.~\ref{figure:histogram}.
Similar to the results of de Jong (\cite{dejong:3}) for early spiral types,
the median disk has a $\mu_d^c(K)$ of 17.1\magarc, $r_d=4.6$ kpc;
with $M_d(K)=-24.3$ mag it is slightly more luminous than
the median bulge.
The median ratio of $r_d$ and isophotal (optical) radius $R_{25}$ is 0.24
(shown in Fig. \ref{figure:sl} as a dotted line in the upper left panel),
comparable to what is found in late-type spirals 
(Giovanardi \& Hunt \cite{giova:1988}, Giovanelli et al. \cite{giovanelli}).
Although similar in size, these early-type disks
are more than 1 $K$-mag~arcsec$^{-2}$ brighter than those
in late-type spirals (Giovanardi \& Hunt \cite{giova:1988}).
 
The right panels of 
Fig. \ref{figure:fp} show scatter plots of disk $\mu_d(K)$ vs. $r_d$
(upper panel) and of disk $<\mu^c(K)>_e$ vs. $r_e$ (lower panel).
As for the bulge, 
correlations of disk $\mu_d^c(K)$ with $r_d$ have been noted for some time
(e.g., Kent \cite{kent:1985}).
It is interesting to note that,
when plotted in terms of the photometric observables commonly used in FP
studies (lower-right panel), disks
dwell in 
a region of this FP projection which is contiguous
and similar in shape and extent to that
of bulges. 
The disks appear to extend the bulge relation to
larger radii and fainter surface brightnesses.
Also evident, in the lower-right panel, is 
the rough consistency with the slope for bulges
found by Andredakis et al. (\cite{andredakis:peletier}),
although with a large offset.
It is clear from Fig. \ref{figure:fp} that, unlike the bulge,
the relation between disk parameters does not vary substantially with 
bulge $n$.

\subsubsection{Disk colors \protect\label{results:disks.colors}}

The median $(J-K)_d$ of 0.94 (0.91 for the np decomposition)
is similar to the central colors of late-type spirals
(Frogel \cite{frogel:1985}; Giovanardi \& Hunt \cite{giova:1988}).  
The scatter about the mean is 0.08~mag, smaller than
for $(J-K)_b$.
The median disk is 0.12~mag bluer than the bulge, an effect
not noted by Terndrup et al. (\cite{terndrup}) whose sample 
was dominated by later types.
As for bulges, redder disks tend to be ``denser'' 
(98\% significance ), and $(J-K)_d$ is independent of $M_d$ and inclination. 

\subsubsection{Opacity of the disks \protect\label{opacity}}

Although
the diagnostics of dust content in galaxies have been extensively revised in
recent years (e.g. Byun et al. \cite{byun} -- BFK; Bianchi et al. 
\cite{bianchi:ferrara}), such studies have made 
clear that disk opacities are not easy to determine. 
In the following we gather the indications about disk opacity obtained here;
none of the tests we adopt is particularly
stringent, due mainly to the small number statistics, but all 
converge on conservative estimates for the opacity of early
type disks: $\tau_V(0)$ ranges from 2 to 4,
where $\tau_V(0)$ is the central optical depth in the $V$ band (face-on).
This is essentially the same result reached by Peletier \&
Willner (\cite{peletier:willner}) and Giovanardi \& Hunt 
(\cite{giova:1996}). 

(a) Correlation between apparent disk brightness and inclination.
We find a slight trend in both bands,
with slopes $C_J = 0.66\,\pm\,0.23$ and 
$C_K = 0.73\,\pm\,0.23$, both compatible with a fully
transparent disk (with $C=1$, see Eq. \ref{eqn:transparency}).
Taken at face value, a $C=0.7$
corresponds to a $\tau_V(0)$ of $\sim$ 1.1 if measured in the $J$ band, and
to $\tau_V(0) \simeq 1.8$ if in $K$\footnote{From Fig. 8 in BFK and assuming 
$A(\lambda)/A(V)$ as in Cardelli et al. (\cite{cardelli});
these estimates have been corrected approximately for the lower dust albedo
in the NIR.}. 
These moderate values for the central opacity
imply that the spread observed in the NIR $\mu_d^c$ is intrinsic and
not due to extinction;
de Jong (\cite{dejong:4}) reached a similar solution 
on the basis of a larger sample.

(b) Variation of disk scale length with wavelength and inclination
(Evans \cite{evans}; Peletier et al. \cite{peletier}).
Five of the sample galaxies
have been parametrically decomposed in the optical: either $r$
(Kent \cite{kent:1985}; NGC 2639), $V$ (Kodaira et al. 
\cite{kodaira:watanabe}; NGC 3898, 4698), or $B$ (Boroson \cite{boroson}; 
NGC 2775, 2841, 3898).
We find a trend in the ratios of our to their $r_d$: for the
only measurement in $r$ the ratio is exactly 1, it decreases to
0.85 in $V$, and to 0.70 in $B$. In addition, these ratios 
depend on the inclination, 
thus providing a test which is largely free from
the influence of intrinsic color gradients. 
The correlation, in the sense
of smaller ratios for higher $i$, implies $\tau_V (0) \leq 3$.
These results are consistent with Peletier et al. (\cite{peletier}),
who find that $B$ and $K$ scale length ratios vary from 1.2 to 2.0,
and with inclination.

(c) Colors.
As noted in the previous section, $(J-K)_d$ does not depend 
on $i$. We estimate the maximum (3 $\sigma$) slope of $(J-K)_d$ vs.
$\sec\, i$ which is still compatible with our data to be 0.075. For a
Triplex model (Disney et al. \cite{disney:davies}) with 
$\zeta = 0.5$ (Peletier \& Willner \cite{peletier:willner}),
this implies a $\tau_V(0) \leq 3$.
We noted in Sect.~\ref{results:disks.colors} that
red disks were associated with bright $\mu_d^c$, which again
points to moderate opacities. Indeed,
since $\mu_d^c$ is corrected for inclination
assuming transparency, a high opacity would translate
into faint brightnesses for reddened disks.

\subsection{Relationship between bulge and disk \protect\label{results:bd}}

The best-$n$ median B/D ratio is 0.8 with values ranging
from 0.2 to 2; even in this early-type sample,
more than two thirds of the galaxies have disks more luminous than bulges.
With the exception of NGC~1024,
B/D ratios obtained from the np decomposition are always less than 1
(as can be seen from the lower left panel in Fig.~\ref{figure:sl}).
The two methods yield B/D values which differ by almost
a factor of $\sim 2$, but with large scatter.
We have verified that this is mainly imputable to the choice of 
$\epsilon_b$ and $i$, the values adopted in the np case being lower.
Our parametric B/D ratios are comparable, although somewhat larger, 
to those (parametric) found by Kent (\cite{kent:1985}) in the $r$ band. 
Also, our B/D's (both parametric and np) in the $K$ band
are 10 $\sim$ 15\% larger than in $J$. 
That the $K$-band B/D is larger than in the optical was also
noted by de Jong (\cite{dejong:3}), but with values 
smaller than ours due to his choice of $n$~=~1 bulges.
It is evident that the B/D ratio is rather model 
dependent, and different decomposition methods provide
estimates differing by factors of 2 or more, as illustrated in  
Fig. \ref{figure:sl} where $\log r_e$, $\log\,r_e/r_d$, and $\log\,B/D$
are shown as a function of optical (isophotal) 
radius\footnote{These were taken from Table \ref{table:sample}.}.
Inspection of this figure also shows that the derived bulge and disk parameters,
including best $n$, are not appreciably affected by biases associated with 
galaxy apparent size.

A linear correlation between $r_e$ and $r_d$ over all spiral types
has been recently found by
de Jong (\cite{dejong:3}) and Courteau et al. (\cite{courteau}).
They interpret the correlation as an indication that the Hubble sequence
is scale-free since the relative size of bulge and disk does not depend
on morphological type.
It can be seen from the middle panels that our data are also consistent with
constant $r_e/r_d$;
the sample median best-$n$ $r_e/r_d$ of 0.3 is shown as a dotted line.
This value is a factor of 2 larger than that found by 
Courteau et al.  in the $r$ band, $\langle r_e/r_d \rangle\,=\,0.13$ 
and by de Jong 
for the $K^\prime$ data alone,  $\langle r_e/r_d \rangle\,=\,0.14$
\footnote{Their values
have been converted to bulge effective scale lengths
using the conversion factor for $n$~=~1 given in the Appendix.}. 
However, $r_e/r_d$ appears to be strongly influenced by the bulge 
parameterization: the dashed lines shown in the middle right
panel of Fig. \ref{figure:sl} illustrate the values obtained from
our $n$~=~1 fits (sample median $r_e/r_d$~=~0.2),
and for the $n$~=~4 fits (median $r_e/r_d$~=~0.7). 
According to whether bulges are fit with a simple exponential or 
with the de Vaucouleurs law, $r_e/r_d$ changes by more than a factor of 3.
Hence, if the best-fit $n$ changes with morphological type as suggested by
Andredakis et al. (\cite{andredakis:peletier}),
the claims made by Courteau et al.
for a scale-free Hubble sequence may be premature.

\subsection{Color gradients \protect\label{results:cgrad}}

Since we adopt the same scale length in $J$ and $K$,
our parametric decompositions yield bulges and disks with uniform color. 
On average, the resulting bulges are redder 
than the disks by more than 0.1 mag, and 
we should detect a significant color gradient 
at the transition between bulge and disk. 
Such gradients are clearly evident in 
NGC~3593, 3898, 4419, 4845, 6314 and IC~724, 
all objects whose bulge and disk colors differ greatly.
When such gradients are present, they also appear, enhanced, 
in the $r-K$ profiles.

\begin{figure}
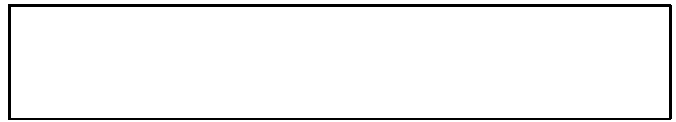

\picplace{1.5cm}
\caption{Colors and color gradients vs. inclination.
The upper panels show
the mean of $J-K$ and $r-K$ colors beyond 3 kpc from the 
center. Color gradients evaluated in 
the same region are shown in the lower panels.}
\label{figure:c_cg}
\end{figure}

Regarding the gradients within the single components, 
we give no estimate of bulge color 
gradients\footnote{This could be accomplished,
in principle, using np profiles or the inner part of the
obseved color profiles but they are difficult to detect.
Due to the seeing, the very inner color profile is hardly reliable,
while at outer radii the bulge brightness rapidly falls deep below the 
disk.}.
For the disk, following Terndrup et al. (\cite{terndrup}), we estimated outer 
($>$ 3 kpc) color gradients, computed
by fitting $J-K$ and $r-K$ versus $\log r$ (in kpc); they will be denoted 
with $\delta(J-K)$ and $\delta(r-K)$ 
respectively\footnote{We used global profiles rather than np 
disk profiles because the
latter are not available for $r-K$, and because the two types of profile are
very similar in $J-K$.}.
We detect $J-K$ gradients at 3\,$\sigma$ in only one case, NGC~6314, 
and in six if we consider a 2 $\sigma$ limit.
All galaxies with $J-K$ gradients for which we have $r$ data,
namely NGC~3593, 3898, 4378, 4419 and 6314, also show 
significant $\delta(r-K)$.
Significant $\delta(r-K)$ is also found in NGC~2639, 4845 and IC~724. 
In agreement with de Jong (\cite{dejong:4}), we only find negative gradients,
ranging from $-0.24$ to $-0.47$ mag per decade in $J-K$, and from $-0.26$ to 
$-1.03$ in $r-K$. 
There is no correlation between $\delta(J-K)$ and inclination, but
inclined galaxies have steeper $r-K$ profiles (see 
Fig.~\ref{figure:c_cg}).

Such color gradients provide a last assessment of the disk opacity,
already discussed in $\S$\,\ref{opacity}.
While NIR colors are stable across the disk, 
we find a prevalence of negative trends in $r-K$,
especially in inclined galaxies (see Fig.~\ref{figure:c_cg}). 
A weighted fit of $\delta(r-K)$ vs. $-2.5\log\, (\cos\, i)$ yields a slope of
$-0.30\,\pm\,0.08$. In $B-I$, a common feature of the models
(e.g. BFK) is that gradients
tend to steepen with increasing $i$ only in rather transparent
disks; for $\tau_V(0) \geq 3$ the trend is reversed due to a saturation 
effect. It has also been shown by Bianchi et al. (\cite{bianchi:ferrara}) 
that disk color gradients are not greatly influenced by the 
dust scattering properties.
In $r-K$ the reddening will be larger by
a factor of 2 for the same $\tau_V(0)$, so the 
observed slope is roughly indicative of a $\tau_V(0)\simeq 1.5$ (BFK; 
Bianchi \cite{bianchi}).

\section{Summary \label{summary}}

We have decomposed $J-$ and $K-$band images of 14 early-type
spirals into bulge and disk components.
2D non-parametric solutions and results from fitting a parametric
model of generalized exponential $1/n$ bulges and simple exponential disks
are compared, and general characteristics of early-type spiral bulges
and disks are examined. 
We find that:
\begin{enumerate}
\item
Even using objective and refined techniques,
the decomposition in structural components is far from being a robust and
unique process.
For the parametric methods,
significantly different decompositions are obtained for different bulge
distribution laws.
Non-parametric techniques, on the other hand, appear to be affected by the
choice of the ellipticities of the components which are difficult to
evaluate objectively.
Of the two components, the bulge is the most subject to errors, since the
inner part is masked by seeing, and the outer regions are buried beneath
the disk.
\item
Bulge structural parameters are strongly influenced by the
form of the function used to derive them.
The same bulge, when fitted with small $n$, appears to be ``denser''
(brighter $\mu_e$), more compact (smaller $r_e$),
and less luminous than when fitted with large $n$.
The dispersion of the fitted parameters also increases with $n$. 
\item
The median early-type bulge has a shape index $n$ between 2 and 3, 
$\mu_e^c(K)$=16.8\magarc , and $r_e$=1.6 kpc.
It is also red, with $(J-K)_b$~=~1.06, and redder bulges tend to be
``denser'', that is with brighter $\mu_e$.
\item
As noted by Kent (\cite {kent:1988}), bulges are rarely spherical.
The median intrinsic ellipticity is 0.34, equivalent to a disk
with 50$^\circ$ declination.
This restricts the applicability of non-parametric techniques
to rather inclined systems and suggests treating with caution
the studies which assume spherical bulges.
\item
The median early-type disk with $\mu_d^c(K)$~=~17.1
is more than 1~\magarc\ brighter than later-type disks, and
bluer than the bulge in $(J-K)$ by more than 0.1~mag.
Disk scale lengths agree fairly well with those found by other at
different wavelengths, and we confirm a tendency for NIR disk
scale lengths to be smaller than those at optical wavelengths
(e.g., Peletier et al. \cite{peletier}).
$r_d/R_{25}$ is approximately constant, 0.24, similar to the value
of 0.25 found for late-type spiral disks
(Giovanardi \& Hunt \cite{giova:1988}; Giovanelli et al. \cite{giovanelli}).
\item
Both bulge and disk surface brightnesses correlate with respective
scale lengths, consistently with the projection of the
fundamental plane for ellipticals and spiral bulges
(e.g., Andredakis et al. \cite{andredakis:peletier}).
We note that uncertainties in the decomposition, especially
in the shape index $n$, strongly influence the
position of a bulge within the FP. 
Disks appear to reside in a region of this FP projection
which is roughly contiguous to that of bulges, 
extending the correlation to larger radii and fainter surface
brightnesses. 
\item
We confirm the tendency
for the ratio of bulge and disk scale lengths $r_e/r_d$ to be constant,
noted by de Jong (\cite{dejong:3}) and Courteau et al. (\cite{courteau}).
However, we find a mean (best-$n$)
value $r_e/r_d$~=~0.3, significantly larger than
the value found by de Jong and Courteau et al.;
our $n$~=~1 value of $r_e/r_d$ of 0.2 agrees roughly with their value
of 0.13--0.14, while our $n$~=~4 value is 0.7, more than a factor of 3 larger.
We attribute such differences to different bulge parameterizations
and caution that if best $n$ varies with morphological type, as suggested
by Andredakis et al. (\cite{andredakis:peletier}), $r_e/r_d$ may not
be constant with morphological type, and thus the Hubble sequence may not
be scale free as proposed by Courteau et al. (\cite{courteau}).
\end{enumerate}

\begin{acknowledgements}
We would like to thank the referee, F. Simien,
for a thorough reading of the manuscript and useful suggestions.
This research was partially funded by ASI Grant 95-RS-120.
\end{acknowledgements}

\appendix
\section{Generalized exponential brightness distributions and conversion
factors}

Assuming circular symmetry, we define a generalized exponential bulge
as a surface brightness distribution:

\begin{equation}
I(r) = I_0 ~exp \left[-\left(\frac{r}{r_0}\right)^{1/n}\right]
\end{equation}

\noindent
where $r$ is the projected radius,
$I_0$ the central brightness, $r_0$ the exponential
folding length, and $n$ an integer index ($n\geq1$).
With the change of variable $x^n=r/r_0$, 
it is easily seen that the total luminosity $L$ is: 

\begin{equation}
L = \int_{0}^{\infty}I(r)~2 \pi r~dr = \pi (2n)!~I_0 r_0^2
\end{equation}

It is customary to introduce instead of $r_0$ an effective radius $r_e$,
which encircles half of the luminosity, and instead of $I_0$ an
effective brightness $I_e = I(r_e)$. We derive in the following
the relations between these quantities in the case of a generalized
exponential. 

The relation defining $r_e$ is:

\begin{equation}
\int_{0}^{r_e}I(r)~2 \pi r~dr = \frac{L}{2}~~ ;
\end{equation}

\noindent
with the above mentioned change of variable, this integral yields:

\begin{equation}
e^{-\alpha_n} \sum_{i=0}^{2n-1} \frac 
{\alpha_n^{2n-1-i}}{(2n-1-i)!} = \frac{1}{2}\, ,\hspace{0.3cm}{\rm with}~~ 
\alpha_n=\left(\frac{r_e}{r_0}\right)^{\frac{1}{n}}, 
\end{equation}

\noindent
which we write as:

\begin{equation}
e^{-\alpha_n} \sum_{i=0}^{m} \frac {\alpha_n^{i}}{i!} = \frac{1}{2}\hspace{1.5cm}{\rm with}~~~~m=2n-1.
\label{eqn:alpha}
\end{equation}

\noindent
The left hand side of this last equation is a cumulative Poisson
distribution with parameter $\alpha_n$. This can be written in terms
of $\chi^2$ probability functions (Abramowitz \& Stegun  \cite{abr:ste}):

\begin{equation}
e^{-\alpha_n} \sum_{i=0}^{m} \frac {\alpha_n^{i}}{i!} = Q(\chi^2|~\nu)\, , 
\end{equation}
with $\chi^2=2 \alpha_n$ and $\nu=2(m+1)=4n$. 

\noindent
$\alpha_n$ is therefore easily evaluated from $\chi^2$ tables. 
In any case, for large $\nu$ values, $(\chi^2/\nu)^{1/3}$ is
approximately normally distributed:

\begin{equation}
Q\left(\chi^2|~\nu\right) \simeq Q(x_2)\, ,
\end{equation}
\begin{equation}
{\rm with} \hspace{0.5cm}x_2=\frac{\left(\frac{\chi^2}{\nu}\right)^{\frac{1}{3}}
-\left(1-\frac{2}{9\nu}\right)}{\sqrt{\frac{2}{9\nu}}}\,.
\end{equation}

\noindent
Since in our case $Q(x_2)=\frac{1}{2}$, 
it follows $x_2=0$ and $\chi^2\simeq\nu(1-\frac{2}{3\nu})$\,. 
Then, with the approximations adopted:

\begin{equation}
\alpha_n\simeq 2n-\frac{1}{3} , ~~~~~~~{\rm or}~~~~~~~r_e\simeq r_0 
\left( 2n-\frac{1}{3}\right)^n
\label{eqn:approx}
\end{equation}

\noindent
It turns out that the approximation is already quite good for $n~=~$1, 
with a relative error of 0.7\%. 
Table \ref{table:appendix} reports the solution of Eq. \ref{eqn:alpha}
and its approximation with Eq. \ref{eqn:approx} for various values of $n$.

For the effective surface brightness we have:
\begin{eqnarray}
I_e &= &I(r_e)=I_0~exp\left[-\left(\frac{r_e}{r_0}\right)^{\frac{1}{n}}\right] \nonumber \\
&=&I_0~e^{-\alpha_n}\simeq I_0~exp~\left(-2n+\frac{1}{3}\right)\, .
\end{eqnarray}

\noindent
These results are easily generalized to elliptical distributions and 
they are found to hold in the same form if $r_e$ is the major
semiaxis of the ellipse encircling half luminosity.
\begin{table}
\begin{flushleft}
\caption{$r_e$ approximation}
\label{table:appendix}\protect
\begin{tabular}{rrrr}
\noalign{\smallskip}
\hline 
\noalign{\smallskip}
n & $\nu$ & $\alpha_n~~~~~$ & $2n-\frac{1}{3}$ \\
\noalign{\smallskip}
\hline
\noalign{\smallskip}
1 & 4 & 1.6784 & 1.6667 \\
2 & 8 & 3.6721 & 3.6667 \\
3 & 12 &5.6702 & 5.6667 \\
4 & 16 & 7.6693 & 7.6667 \\
6 & 24 & 11.6684 & 11.6667 \\
10 & 40 & 19.6677 & 19.6667 \\
\noalign{\smallskip}
\hline 
\end{tabular}
\end{flushleft}
\end{table}

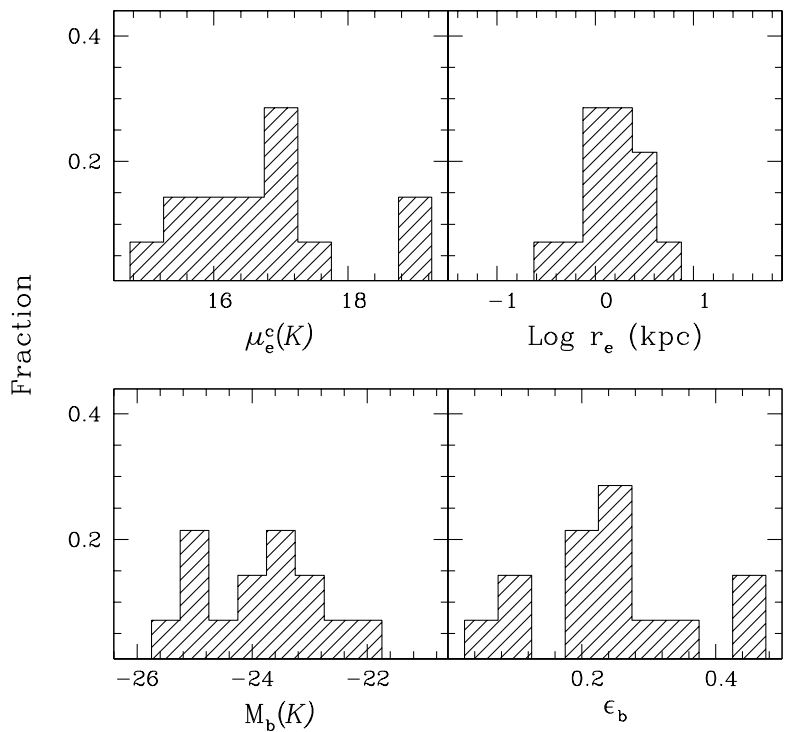
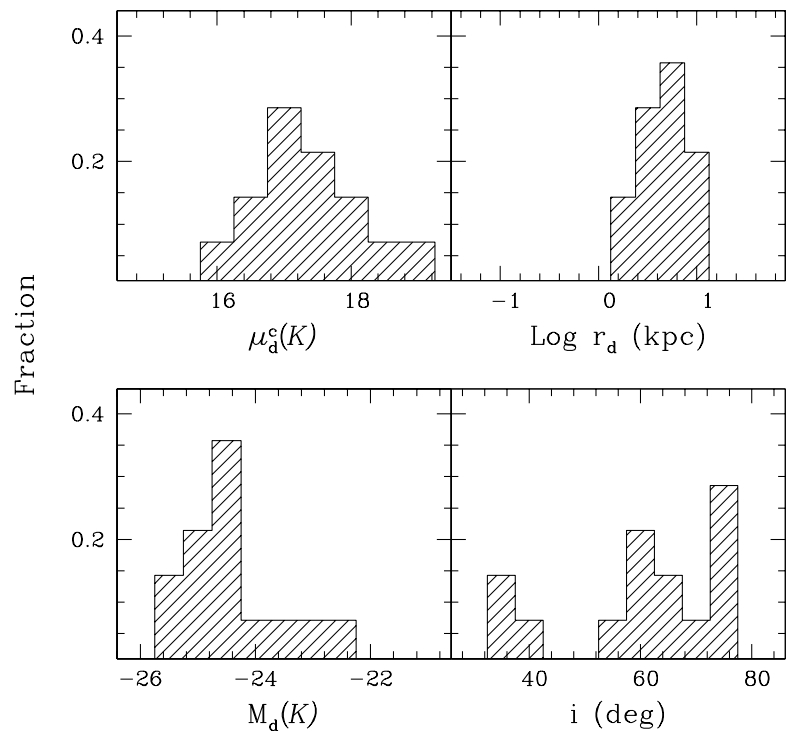
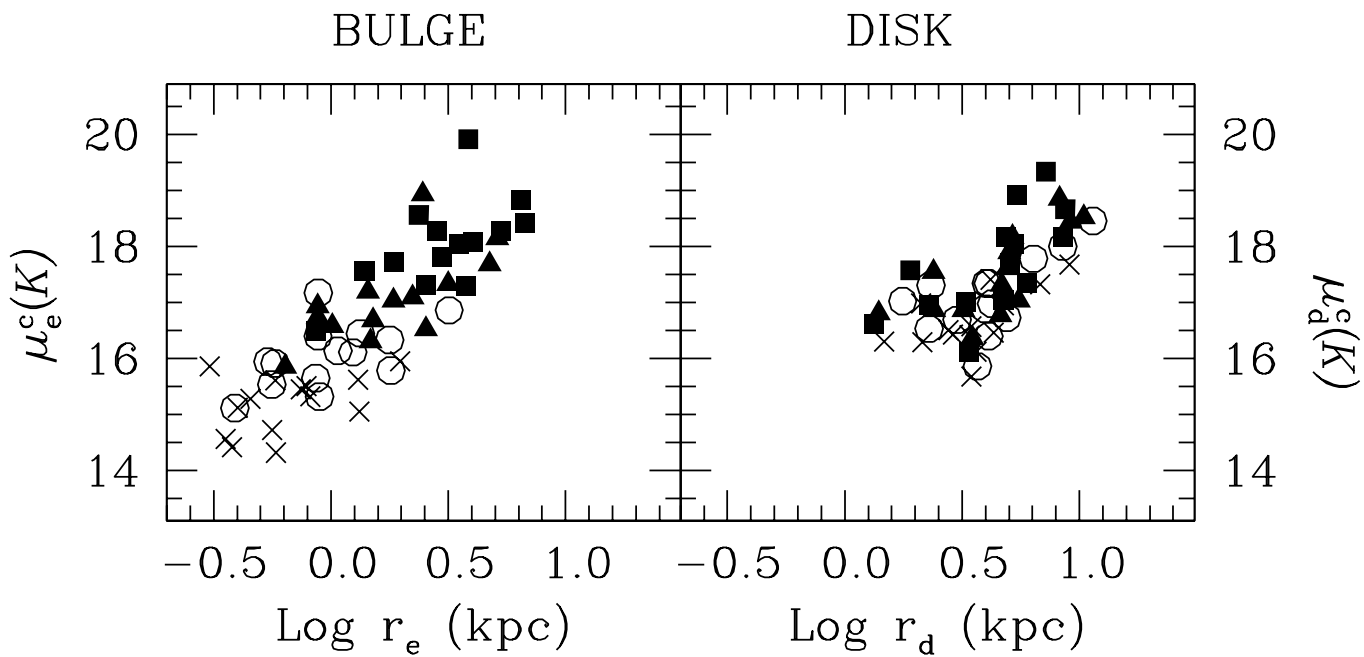
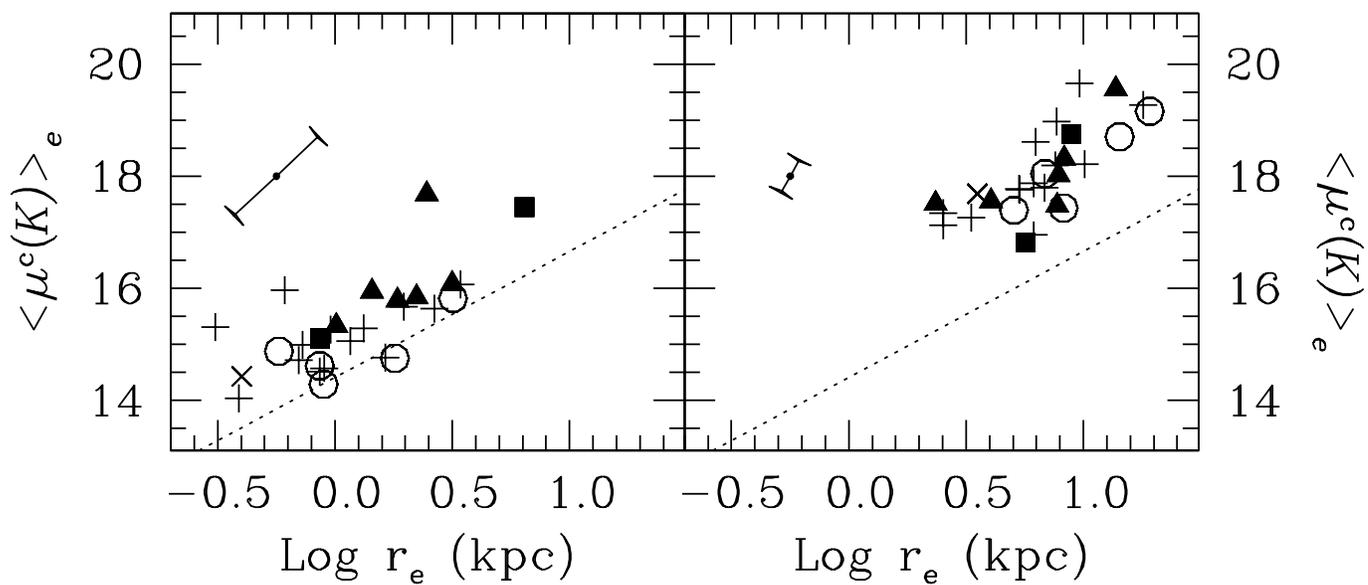
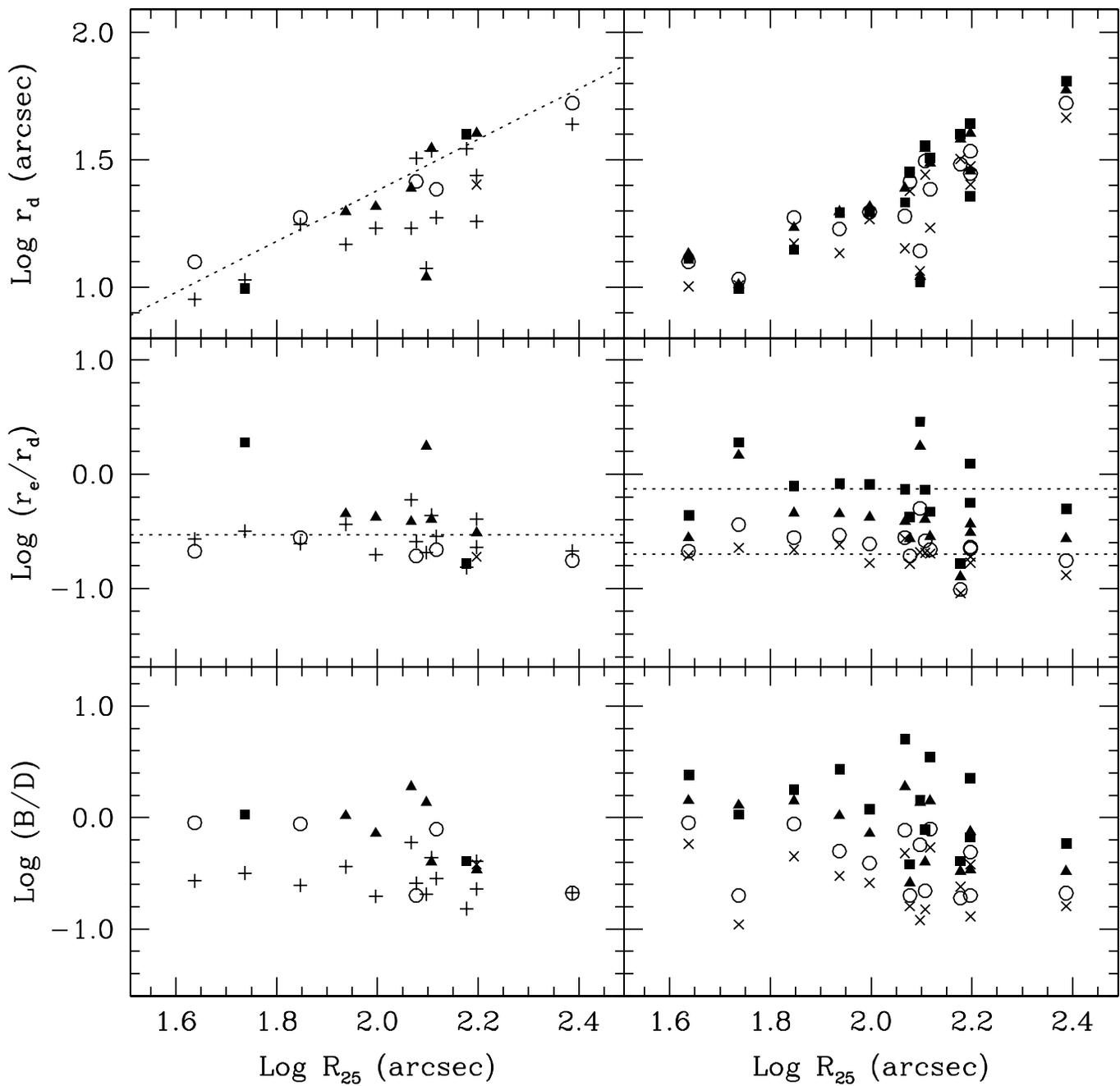
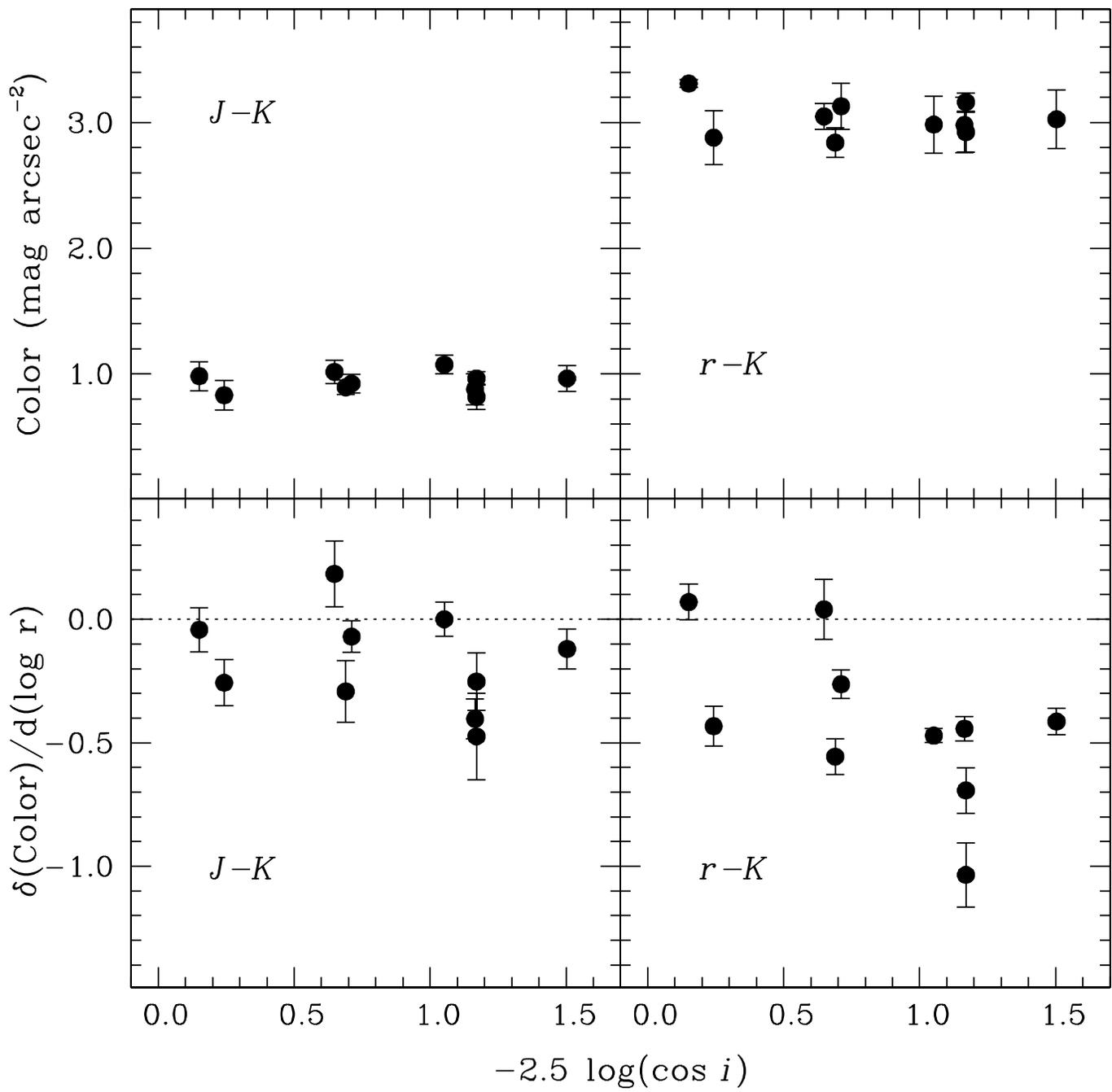
\end{document}